\documentclass[showpacs,preprintnumbers,amsmath,amssymb,showkeys]{revtex4}

\usepackage{graphics,epsfig}
\usepackage{graphicx}
\usepackage{epstopdf}
\usepackage{amssymb}
\usepackage{amsmath}
\usepackage{amsfonts}
\usepackage{bm}
\usepackage{natbib}

\begin{document}

\title{A Lorentz-Poincar{\'e} type interpretation \\
of the Weak Equivalence Principle.}

\date{\today}
\author{Jan (B.) Broekaert} \email{Jan.Broekaert@vub.ac.be}\affiliation{Philosophy of Physics Group, \\ University of Oxford (temp.\-affil.)\\  \& CLEA-FUND,\\ Vrije Universiteit Brussel }

\begin{abstract}
\noindent
The  validity of the Weak Equivalence Principle relative to a local inertial frame is detailed in a scalar-vector gravitation model  with Lorentz-Poincar\'e type interpretation.  Given the previously established first Post-Newtonian concordance of dynamics with General Relativity,  the principle is to this order compatible with GRT.   The gravitationally modified Lorentz transformations, on which the observations  in physical coordinates depend,  are shown to provide a physical interpretation of \emph{parallel transport}.   A development of  ``geodesic'' deviation in terms of the present model is given as well.
\end{abstract}

\pacs{04.20.-q, 04.50.+h}
\keywords{physical relativity, equivalence principle}

\maketitle

\section{Introduction} 
Viable formulations of gravitation should  fulfill, al least to the required experimental accuracy, the Equivalence Principle (EP) \citep{Will1993} (see  \citet{Damour2001} for comments). The Weak Equivalence Principle (WEP)  purports the indistinguishability of inertial and gravitational mass.   The EP on the other hand requires physical laws to be equivalent  in all \emph{local} free-falling  frames and, equivalent with their expression in unaccelerated frames  without a gravity field.  The WEP can also be stated as the principle of universality of free-fall or, that gravitation vanishes for the free-falling observer.  It is well known however that such an observer will only locally establish that result. Each small spatial separation between a free-falling observer and some free-falling point particle  ---even if at some instance it was at rest relative to the falling observer--- will cause a relative acceleration, e.g.  \citet{Weinberg1972}, p. 148.  
The free-falling observer must thus be reduced to a \emph{local} inertial frame (LIF). \\
A free-falling observer   \emph{crossed}  by some free-falling system with a \emph{non-zero} relative velocity, must ---according to the WEP--- have a zero relative acceleration only while meeting at the intersection point of their trajectories. That this is the case for a LIF observer is \emph{a priori} not intuitive given the free-fall acceleration relative to a \emph{static} observer \citep{McGruder1982,BunchaftCarneiro1998,Broekaert2005a}. 
Damour's stipulation of the Weak Equivalence Principle, or ``universality of free-fall", is precise in this sense  (\citep{Damour1994}, emphasis added):
{\quote \small C3: ÒPrinciple of geodesicsÓ and universality of free-fall : small, electrically neutral, non self-gravitating bodies follow geodesics of the external spacetime (V, g). In particular, two test bodies dropped at the same
location and \emph{with the same velocity} in an external gravitational field fall in the same way, independently of their masses and compositions. \hfill (a)}\\

\noindent In GRT, a \emph{static} observer  will attribute  a free-fall acceleration with explicit dependence  on the kinematics of a particle --- using generic coordinates  \citep{BunchaftCarneiro1998, Bini1995,McGruder1982} (a coordinate-free space-time decomposition of a covariant expression):
\begin{eqnarray}
\mathbf{a}  & = &  \mathbf{g} -   \mathbf{v}  ( \mathbf{v}.\mathbf{g}) 
\end{eqnarray}
where $\mathbf{a}$ is the general  local 3-proper-acceleration of a  particle  in a static gravitational field  and $\mathbf{g}$  is this same acceleration but with the ``physical" relative  velocity  $\mathbf{v} = 0$.  Following the choice of adapted coordinates (e.g. Fermi coordinates) this equation amounts to a kinematic decomposition of the observed gravitational acceleration, it is not a transformation.
From this decomposition however it follows that  two locally coincident  free-falling point particles will  expose  a \emph{mutual} relative acceleration  to  the static observer ---the latter third object acting as the reference frame. This mutual acceleration is to be understood as the  acceleration of the connecting vector of the two bodies in terms of the difference of their location as measured by the static observer, e.g. \citet{Rindler1979}, p. 36.
For a \emph{static} observer; at the coincidence of two free-falling particles the observed mutual acceleration does in general not vanish;  a term due to kinematical differences between the particles remains:
\begin{eqnarray}
\mathbf{a}_{rel._{12}} \ \equiv \ \mathbf{a}_1 -  \mathbf{a}_2 &=& -   (\mathbf{v}_1.\mathbf{g}) \mathbf{v}_1  +   (\mathbf{v}_2.\mathbf{g}) \mathbf{v}_2  \label{StatRelAccDiff}
\end{eqnarray}
This kinematical feature ---apparent to a static observer--- is still conform with the WEP statement (a) by Damour. In previous work we have shown that in the Lorentz-Poincar\'e model ---introduced in the next section--- the same kinematical effects are apparent in the static observer case. In the latter model the acceleration was obtained by a coordinate transformation; from coordinate space and time to local coordinates of the static observer \citep{Broekaert2005a}, while the coordinate space acceleration itself was obtained by a Hamiltonian principle \citep{Broekaert2005b}. \\
In order now to validate the WEP such that  ``the equivalence of acceleration and gravitation is realized" (b) ---in comparison to (a)---  the local dynamics of one or more non-mutually interacting particles \emph{related to a free-falling frame} should be that of  free particles and  no kinematical relative or mutual accelerations should remain at coincidence.  The instant free-fall of the previously static observer ---mentioned above--- should indiscriminately annihilate all these residual relative kinematical accelerations (\ref{StatRelAccDiff}) between random free-falling particles. \\
In GRT for a Local Inertial Frame it is well known that these accelerations are zero.  The covariant derivative of a tensor ---e.g. the four-velocity---  can be expressed as the sum of the ordinary derivative and changes of the tensor due to \emph{parallel transport} (e.g. \citet{Kenyon1990}, sec 6.1, or \citet{Stephani2004}, sec 18.3);
\begin{eqnarray}
\frac{D U^{a}}{D \lambda} & = & \frac{d U^{a}}{d \lambda} + \Gamma^{a}_{nm} U^m \frac{d x^n}{d \lambda} \label{covderiv}
\end{eqnarray}
The connections thus express the change of a tensor along a certain spacetime curve. In the LIF  coordinates the connections are locally zero and any four-acceleration of a free-falling particle crossing the LIF-observer will be momentaneously zero: in the LIF's local coordinates the four-velocity does not vary during the (infinitesimal!) parallel transport while in coordinate space the four-velocity does vary according to  the equation of motion of the particle.\\
The vanishing of the relative accelerations (\ref{StatRelAccDiff}) for spatially coinciding systems is thus realized by having recourse to the  \emph{covariant} derivative, Eq. (\ref{covderiv}),  in the definition of the particle's acceleration. In the coordinate space description the \emph{zero} covariant acceleration is invoked to obtain the (non zero) particle acceleration $d \mathbf{U}/d \lambda$ over its trajectory (e.g. \citet{Weinberg1972}  p 212), while  in local coordinates of the LIF the zero covariant acceleration reduces to a zero physical relative acceleration due to the vanishing of the connections.\\
 Now in the L-P model as well we will require the introduction of  \emph{parallel transport} in the definition of the derivative. We will detail below how this procedure naturally emerges in the L-P model and, leads in the model's context to the validity of the WEP in the LIF perspective as well. \\
A short introduction to the L-P model is given in the following section, a detailed development and calculation can be found in  our previous work \citep{Broekaert2005a,Broekaert2005b,Broekaert2004b,Broekaert2002}. We note that Lorentz-Poincar\'e type properties have been studied as well in alternative models of gravitation in the literature; e.g. the validity of the WEP in a  \emph{scalar} gravity model ---concerning the point-particle limit of an extended body--- was recently discussed in terms of an ``unaffected Euclidean metric" by \citet{Arminjon2006}.

\section{Gravitation model with Lorentz-Poincar\'e type interpretation}
The  L-P gravitational model maintains the effects of  length  shortening (``rod contraction") and time dilation (``clock slowing") as we understand them in the Lorentz-Poincar\'e interpretation of Special Relativity. A recent discussion of  this  ``dynamical" interpretation of SRT based on the Lorentz-covariance of the fundamental interaction of the micro constituents in rods and clocks is given in  \citet{Brown2005}, see also \citet{Bell1987}.  Notwithstanding that Lorentz's electromagnetic preliminary gravitation theory \cite{Lorentz1900} nor Poincar\'e's Newtonian Lorentz-covariant gravitation theory \cite{Poincare1906} were not conceived along the line here presented; we introduce the term L-P \emph{type} because of the continuity with the Lorentz-Poincar\'e interpretation of SRT originally pertaining to physical effects on configurations of matter due to motion.
Now however, these effects are due both to  position in the gravitation field  as well as a  relative kinematics \citep{Broekaert2004b,Broekaert2005b}.   This L-P type of development will explicitly use two levels of description: gravitationally affected observations \emph{versus} gravitationally unaffected ``observations". 
A similar procedure is  used in field approaches to relativity where ``unrenormalized" and ``renormalized'' coordinates are distinguished, see e.g. the work of \citet{CavalleriSpinelli1980, Thirring1961, Dehnen1960, Dicke1957, Wilson1921} and, compare to \citet{Brown2005} for a similar dynamical analysis of GRT  sans issue of gravitational effects on observations, and \citet{Dieks1987} and \citet{Sexl1970}  for the relation with geometric conventionalism.\\
Note that, as the unaffected perspective corresponds to the \emph{coordinate} space description in GRT, it can not truly be considered \emph{observable}. Moreover gravitation can not be shielded from, thus at most can the unaffected perspective be calculated starting from observable affected quantities.  Similarly in GRT, e.g. \citet{Rindler1979} p 142,  a transformation of \emph{coordinate} time and distance into \emph{local} time and distance is required in order to obtain observable quantities.  \\
The gravitational effects on space and time observations were developed as a gravitationally modified  Lorentz Transformation (GMLT) for space and time intervals. In particular these transformations  relate affected and unaffected descriptions. It was also shown that  the elimination of the unaffected perspective  from the GMLT between two local observers restores the local Lorentz covariance of the relations \citep{Broekaert2005a}. Therefore ---even as the GMLT expose the spatial variability of the velocity of light in coordinate perspective--- the \emph{locally observed} velocity of light, $c'$, remains the universal vacuum value.\\  In this model, related but distinct GMLT's for energy and momentum were fitted to the static Newtonian potential energy.  These  transformations give the Hamiltonian expressions  for particles and photons in the unaffected perspective by simply assuming the special relativistic expressions in the affected perspective. With the resulting Hamiltonian, the equations of motion  verify till 1-PN the gravitational phenomenology of GRT \citep{Broekaert2005b}.   In fact, in the L-P model each quantity with different physical dimension is expected to transform according to  a different power of the scaling function and according to  covariant or contravariant GMLT's, the former aspect is similar to the gravitation model by \citet{Dicke1957}.\\
We state explicitly now the space and time GMLT for further developments in the next section. Let a  physical ---thus affected--- observer at coordinate position $\mathbf{r}$ locally measure space and time intervals $(d\mathbf{x}', dt')$. The space and time GMLT ---for which we will adopt the standard mathematical symbol for the Lorentz transformation $\Lambda^\mu_{\ \nu}( \mathbf{v}, \mathbf{r})$ but now with two arguments; velocity and space (and time) location--- will relate these to  intervals $(d\mathbf{x}, dt)$ in the unaffected perspective \citep{Broekaert2005b}:
\begin{eqnarray}
\left(\begin{array}{c} dt' \\ d{\bf x}' \end{array} \right)
\ = \ \Lambda(\mathbf{u}, \mathbf{r})
\left(\begin{array}{c} dt \\ d{\bf x} \end{array} \right) ,    && \Lambda(\mathbf{u}, \mathbf{r}) \ \equiv \ \left(\begin{array}{ll}
 \gamma \Phi  &  -\mathbf{u} c^{-2} \gamma \Phi \\
 -\mathbf{u}  \gamma \Phi^{-1}  &  \mathbf{1}  \Phi^{-1} + \frac{\mathbf{u}_i\mathbf{u}_j}{u^2} (\gamma-1)  \Phi^{-1} \end{array} \right) \label{GMLT}
\end{eqnarray}
where $\Phi = \Phi (\mathbf{r})$, $c= c' \Phi^2$ and $\gamma = (1- u^2/c^2)^{-1/2}$. We remark that  the inverse GMLT ---transforming $S'$ into $S_0$ quantities---  is given by:
\begin{eqnarray}
 \Lambda^{-1}(\mathbf{u}, \mathbf{r}) &\equiv& \left(\begin{array}{ll}
 \gamma \Phi^{-1}  &  -\mathbf{u}' {c'}^{-2} \gamma \Phi^{-1} \\
 -\mathbf{u}'  \gamma \Phi   &  \mathbf{1}  \Phi  + \frac{\mathbf{u}'_i\mathbf{u}'_j}{{u'}^2} (\gamma-1)  \Phi  \end{array} \right) \ = \ \left(\begin{array}{ll}
 \gamma \Phi^{-1}  &  \mathbf{u} {c}^{-2} \gamma \Phi \\
 \mathbf{u} \gamma \Phi^{-1}  &  \mathbf{1}  \Phi + \frac{\mathbf{u}_i\mathbf{u}_j}{{u}^2} (\gamma-1)  \Phi  \end{array} \right) \label{InverseGMLT}
\end{eqnarray}
The second member is written in hybrid form ---the expression contains $S_0$ terms;  $\mathbf{u}$ and $c$ instead of $\mathbf{u}'$ and $c'$---   being better adapted to use in the next section.\\
In the case of a \emph{non-stationary} source, the GMLT must take into account the \emph{induced velocity} field $\mathbf{w}$ caused by source movement \citep{Broekaert2004b}:
\begin{eqnarray}
\Lambda(\mathbf{u}_0, \mathbf{r}) \ = \ \left(\begin{array}{ll}
 \gamma_0 \Phi  &  -\mathbf{u}_0 c_0^{-2} \gamma_0 \Phi \\
 -\mathbf{u}_0  \gamma_0 \Phi^{-1}  &  \mathbf{1}  \Phi^{-1} + \frac{{\mathbf{u}_0}_i {\mathbf{u}_0}_j}{u_0^2} (\gamma_0-1)  \Phi^{-1} \end{array} \right)
\end{eqnarray}
 by an additional  ``Galilean" relation in coordinate space, according to  a local translation by the field $\mathbf{w}$. The Galilean composition of the effective velocity is considered a first-order approximation of the physical result of the induced velocity field on the classical velocity:
\begin{eqnarray}
d\mathbf{x}_0 =  d\mathbf{x}_w - \mathbf{w} dt_w  &, &     dt_0 = dt_w
\end{eqnarray}
 In GRT  the  quantity corresponding to $\mathbf{w}$ is the `vector potential' $\mathbf{\zeta}$ ---the first relevant order of $g^{i0}$--- caused by the movement of the source in coordinate space; \citet{Weinberg1972}, eq 9.1.62. The frame velocity  $\mathbf{u}_0$  and the velocity of light $ c_0 $ are given by $\mathbf{u}_0 = \mathbf{u} - \mathbf{w}$ and $ c_0 =  \vert \mathbf{c}_w - \mathbf{w} \vert$. We emphasize that while the model deploys spatially-variable speed of light (VSL), the locally observed velocity of light remains the universal vacuum value $c'$ in conformity with the local Minkowski metric. \\
For completeness we note that the gravitational scaling and induced velocity fields $\{\Phi, \mathbf{w}\}$ are given by the equations:
\begin{eqnarray} 
  \Delta  \Phi  \ = \ \frac{4 \pi G' }{{c '}^2} \rho({\bf r} ) \Phi  +  \frac{\left(\nabla \Phi \right)^2}{\Phi}     &,& \Delta {\mathbf  w} \ = \  -   \frac{16 \pi  G'}{{c'}^2}  \rho {\mathbf v}_\rho ({\mathbf x}, t) 
\end{eqnarray}
in no-retardation approximation \citep{Broekaert2004b}. The L-P model thus relies on a scalar-vector field representation of gravitation; a formalism with historic precedence  but  also recent development, e.g. \citet{Winterberg1988a} and \citet{Vlasov1995}.\\

\section{Acceleration transformations in the L-P model.}
We have shown in previous work \cite{Broekaert2005b,Broekaert2004b} that based on a Hamiltonian description, the L-P model gives explicitly the particle and photon 1-PN gravitational accelerations in the unaffected perspective,  e.g. \citep{Weinberg1972}  Eq 9.2.1 (static field);
\begin{eqnarray}
\mathbf{a} (\mathbf{r}, \tilde{\mathbf{u}})  &=& -({c'}^2 + v^2) \nabla (\varphi + 2 \varphi^2)  + 4 \mathbf{v} \mathbf{v}.\nabla \varphi \label{StaticAcc}
\end{eqnarray}
As we expect,  the basic premiss of the WEP as ``equivalence of gravitation and acceleration" is already valid in the unaffected perspective; the acceleration is independent of the mass and energy of the falling entity. The transformation of this expression into the affected LIF perspective will not impair that quality. The present issue is however how to do this  transformation to the LIF and show that spurious kinematic terms of type Eq. (\ref{StatRelAccDiff}) vanish.\\
 In previous work  \citep{Broekaert2005a}, we found that an acceleration transformation ---as standard time-derivative of the velocity transformation--- in the case of  a \emph{fixed} observer leads to a correct rendition of  GRT relation in similar conditions ($\mathbf{u} = 0$, $\mathbf{w} = 0$) \citep{BunchaftCarneiro1998, Bini1995,McGruder1982}:
\begin{eqnarray}
{\mathbf{a}'} &=&  \Phi^{-3} \left\{    {\mathbf{a} }   - 2{\mathbf{v}' } {\Phi}^2 \dot{\varphi} \right\}\label{avectoavecacstatic}
\end{eqnarray}
 The  acceleration transformation (\ref{avectoavecacstatic})  can not  be adapted to the LIF observer.  The observer in free-fall can not simply observe the  value of a velocity of a remote system.  In the case of the static observer   this  can done by rescaling, in its fixed frame, with respect to the value  of $\Phi$ at the remote location. In the LIF case the scaling of the observer frame itself is changing due to its proper free-fall trajectory as well.\\ 
In the case of the free-falling observer we must take into account  the proper movement of the observer ---frame velocity $\mathbf{u}$ and frame acceleration  $\mathbf{a}(\mathbf{u})$--- and find a procedure to relate  \emph{remote} values of the  velocity  of the observed particle to  \emph{local} values at the final location and time of the observer. The relative acceleration is then defined using the ---both local---  final and intial value of the velocity over a time-interval $dt'$;
\begin{eqnarray}
\mathbf{a}'_{(PT)}  &\equiv& \lim_{dt' \to 0} \left({{{\mathbf{v}'}_f}_\vert}_{local}- {{{\mathbf{v}'}_i}_\vert}_{local} \right)/dt' \label{protorelacc}
\end{eqnarray} 
This reduction to local values of quantities will be done according to  a procedure that amounts to parallel transport in GRT (e.g. \cite{Kenyon1990}, sec. 6.1).

\subsection{Parallel transport \label{subsPT}}
An unequivocal physical procedure for a free-falling observer  is constructed for obtaining a local value from a  remote observation in the gravitational field.
 The procedure requires the aid of an \emph{auxiliary} affected observer: a \emph{free-falling} auxiliary observer on a trajectory such that it evolves from the initial remote location at instance $\mathbf{1}$ to the final local location  at instance $\mathbf{2}$ of the LIF observer.  
The free-fall evolution will  define in an at least locally unique way  the trajectory $\mathbf{1}$ to $\mathbf{2}$, since the \emph{geodesic} corresponds to the longest ---and locally unique--- curve between two spacetime points (e.g. \citet{Stephani2004}, p 99, 131). 
This auxiliary  observer   will consider invariant a previously measured quantity which it ``carries" subsequently along, as we will see in the next subsection. However, again  from the unaffected perspective one would still  consider  that both the observed quantity and the observer's measurement standards are equally  affected. Subsequently the free-falling auxiliary observer  during its evolution retains constant the measured values, because  its measurement standards should appear invariably self-similar as we will demonstrate.  Note thus, that the gravitational and kinematic  effect ---monitored by the GMLT--- continuously vary over the trajectory, while the quantity  measured by the free-falling affected observer remains the same from  the initial till the final instance of the trajectory. This reflects precisely what happens in GRT; from Eq. (\ref{covderiv}) we see that the connection term describes the change in coordinate perspective, while in the LIF ---both of the observer and the transporting frame--- the quantity will remain invariant because the connections vanish.
 From the GMLT at the initial and the GMLT at the final instance the invariant quantities of the affected free-falling auxiliary observer can be eliminated and, a relation between  \emph{coordinate} space  and time intervals  locally and at the remote location is established.  The  procedure ---here developed by GMLT--- precisely expresses parallel transport in GRT  since the relation between the initial point and final point is the \emph{free-fall} evolution, i.e. the geodesic. The connections governing the parallel transport in GRT are implicitly present in the L-P model  in their double contraction over the 4-velocity in the free-fall acceleration,  Eq. (\ref{StaticAcc}), of the transporting frame, e.g. \cite{Weinberg1972}, p 212.   \\
 The transport procedure is formalized subsequently.  We first notice that the tensorial rank of an observed quantity $T$ will determine the precise transformation that relates the affected observer and the unaffected description; i.e. $\Lambda^\mu_{\ \sigma} \Lambda^\nu_{\ \tau} T^{\sigma\tau}$ for rank-2,   $\Lambda^\mu_{\ \sigma}  T^{\sigma}$ for rank-1, etc. Since  Eq. (\ref{protorelacc}) requires the transformation of velocities, the GMLT  (\ref{GMLT}) applicable to the 4-vector of space and time intervals will be used. \\
 At initial space and time  instance $\mathbf{1}$  the $S_0$ space and time intervals $\{dt_1, d\mathbf{x}_1\}$ are transformed to a free-falling observer $S'_{PT}$, the auxiliary transporting observer.  $S'_{PT}$ has velocity $\tilde{\mathbf{u}}_1$ such as to evolve by free-fall to space and time instance $\mathbf{2}$. The  affected intervals observed by $S'_{PT}$ are given by $\Lambda(\tilde{\mathbf{u}}_1,1) (dt_1, d\mathbf{x}_1)$.   During free-fall these affected quantities should remain invariant to $S'_{PT}$. At final instance $\mathbf{2}$ the inverse transformation $\Lambda^{-1}(\tilde{\mathbf{u}}_2,2)$ gives  the \emph{transported} intervals in coordinate perspective of $S_0$ at instance $\mathbf{2}$:
\begin{eqnarray}
\left( \begin{array}{c}
dt_{1_{PT_{12}}} \\ d \mathbf{x}_{1_{PT_{12}}} \end{array}\right) &=& \Lambda^{-1}(\tilde{\mathbf{u}}_2,2)   \Lambda(\tilde{\mathbf{u}}_1,1) \left( \begin{array}{c}
dt_{1} \\ d \mathbf{x}_{1} \end{array}\right) , \  \tilde{\mathbf{u}}_2 = \tilde{\mathbf{u}}_1 + \int^2_1 \mathbf{a} (\mathbf{r}, \tilde{\mathbf{u}}) dt \label{paralleltransport}
\end{eqnarray}
where $\mathbf{a} (\mathbf{r}, \tilde{\mathbf{u}})$ is the free-fall acceleration in coordinate space, which to first Post-Newtonian is given by Eq. (\ref{StaticAcc}).
The application of the transport procedure to the definition of acceleration  Eq. (\ref{protorelacc}) in LIF-perspective is now straightforward. 

\subsection{Acceleration in  LIF-coordinates}
Let a free-falling particle be observed by a  free-falling observer. The LIF-observer attributes an acceleration to the particle, which according to  the WEP should be zero when the particle is spatially coincident with the observer. In order to calculate the acceleration the observer requires the initial and final velocity over an infinitesimal time interval. Let the observer measure the final ---local--- value of the particle's velocity at the intersection of their free-fall trajectories. According Eq. (\ref{protorelacc}) then the initial velocity, an instance $dt$ prior to intersection, should be rendered local to the intersection instance by parallel transport.\\

\hskip 0 cm \includegraphics[width=80mm, height = 50mm]{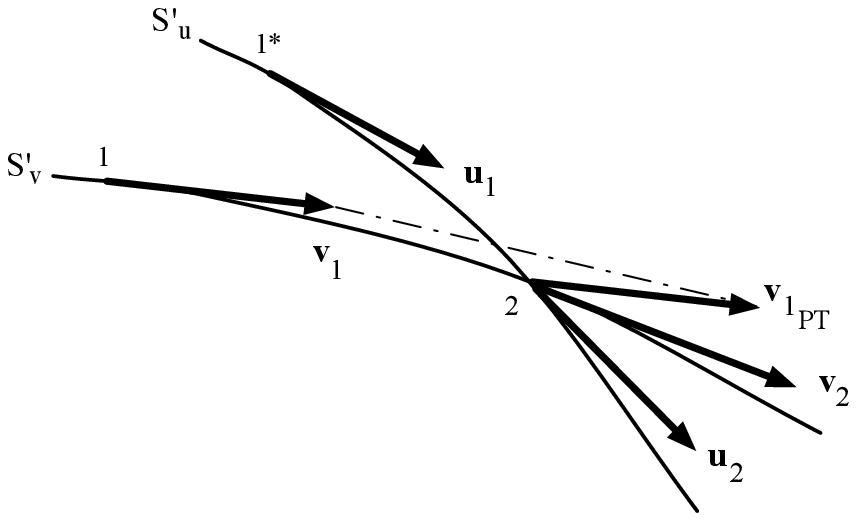}

\vglue - 4.0 true cm
\hangindent = 10 true cm \hangafter = - 30 {\small \em \noindent  Fig.1:  A  scheme of two free-fall trajectories intersecting at location and coincidence time instance $\mathbf{2}$.  An instance $dt$ prior to intersection the observer and the observed particle  respectively had velocities $\mathbf{u}_1$  and $\mathbf{v}_1$. The free-falling particle frame  $S'_v$ effectively  parallel transports the remote initial value  $\mathbf{v}_1$ at $\mathbf{1}$ to the observer $S'_u$ at $\mathbf{2}$ in its form ${\mathbf{v}_1}_{PT}$.  ($\mathbf{1}$ and $\mathbf{1}^*$ are at same time instance)
} 

\vskip 1.5 cm

\noindent 
In the present case (see Fig. 1) the auxiliary transporting frame is identical to the particle's rest frame $S'_v$; from $\mathbf{1}$ it reaches $\mathbf{2}$ at time $t_1+dt$ by free-fall ($t_1=t_{1^\star}$).
The observer frame $S'_u$ evolves from initial location $\mathbf{1}^\star$ to  $\mathbf{2}$ while the particle $S'_v$ parallel transports its initial values from $\mathbf{1}$ to  $\mathbf{2}$.  At the intersection of  the free-fall trajectories $\mathbf{2}$,  $S'_v$ disposes of the required initial velocity $\mathbf{v}_{1_{PT}}$ and final velocity  $\mathbf{v}_{2}$. The parallel transport  is thus described by relation (\ref{paralleltransport}) with initial velocity and final velocity: 
\begin{eqnarray}
 \tilde{\mathbf{u}}_1 \ \equiv \ \mathbf{v}_1 &,& \tilde{\mathbf{u}}_2 \ \equiv  \ \mathbf{v}_2 \ = \ \mathbf{v}_1 + \mathbf{a} (1, \mathbf{v}_1) dt  \label{freefallvelocities}
\end{eqnarray}
where $\mathbf{a}$ is the free-fall acceleration Eq (\ref{StaticAcc}).\\
We want to check the WEP in the affected perspective; we must therefore express the transported quantities $\{dt_{1_{PT}}, d \mathbf{x}_{1_{PT}} \}$ in terms of measurements of  the LIF-observer $S'_v$, using relation (\ref{GMLT}). Then the attributed intervals in affected  perspective are; 
\begin{eqnarray}
\left( \begin{array}{c}
dt'_{1_{PT_{12}}} \\ d \mathbf{x'}_{1_{PT_{12}}} \end{array}\right) &=& \Lambda(u_2,2) \Lambda^{-1}(\tilde{u}_2,2)   \Lambda(\tilde{u}_1,1) \left( \begin{array}{c}
dt_{1} \\ d \mathbf{x}_{1} \end{array}\right)
\end{eqnarray}
The relative acceleration of the particle relative to $S'_v$, according to  (\ref{protorelacc}), requires $dt'$ the  time lapse in affected perspective between instances  $\mathbf{1}$ and $\mathbf{2}$.  Thus  while $dt_1$ and $dt_2$ are the intervals in which the velocities are measured, $dt$ is the interval in which the acceleration is measured. In the unaffected perspective these intervals are all taken identical, while in affected perspective these are given by; 
\begin{eqnarray}
dt'_2 &=& \Lambda(u_2, 2)^0_{\ \mu} ( dt_2,   d \mathbf{x}_2 )^\mu \\
dt'_{1_{PT_{12}}} &=& \Lambda(u_2, 2)^0_{\ \mu}   \Lambda^{-1}(\tilde{u}_2, 2)^\mu_{\ \kappa}   \Lambda(\tilde{u}_1, 1)^{\kappa}_{\ \mu} ( dt_1,   d \mathbf{x}_1 )^\mu \\
dt' &=& \Lambda(u_2, 2)^0_{\ \mu} ( dt,   d \mathbf{x} )^\mu
\end{eqnarray}
Then following Eq. (\ref{protorelacc}) the acceleration is given by
\begin{eqnarray}
{\mathbf{a}'}^k_{(PT)}  &=& 
 \lim_{dt' \to 0}
\frac{\Lambda(u_2, 2)^k_{\ \mu}}{\Lambda(u_2, 2)^0_{\ \tau}}  \left( \frac{( dt,   d \mathbf{x}_2 )^\mu}{( dt,   d \mathbf{x}_2 )^\tau } - \frac{\Lambda^{-1}(\tilde{u}_2, 2)^\mu_{\ \nu}   \Lambda(\tilde{u}_1, 1)^\nu_{\ \sigma}}{\Lambda^{-1}(\tilde{u}_2, 2)^\tau_{\ \rho}   \Lambda(\tilde{u}_1, 1)^\rho_{\ \omega}}  \frac{( dt,   d \mathbf{x}_1 )^\sigma}{( dt,   d \mathbf{x}_1 )^\omega } \right)  \frac{1}{dt'} 
\end{eqnarray}
or in terms of velocities;
\begin{eqnarray}
{\mathbf{a}'}^k_{(PT)}   &=& \lim_{dt' \to 0}
\frac{\Lambda(u_2, 2)^k_{\ \mu}}{\Lambda(u_2, 2)^0_{\ \tau}}  \left( \frac{( 1, \mathbf{v}_2 )^\mu}{( 1,    \mathbf{v}_2 )^\tau } - \frac{\Lambda^{-1}(\tilde{u}_2, 2)^\mu_{\ \nu}   \Lambda(\tilde{u}_1, 1)^\nu_{\ \sigma}}{\Lambda^{-1}(\tilde{u}_2, 2)^\tau_{\ \rho}   \Lambda(\tilde{u}_1, 1)^\rho_{\ \omega}}  \frac{( 1,   \mathbf{v}_1 )^\sigma}{( 1,   \mathbf{v}_1 )^\omega } \right)  \frac{1}{dt'} \label{LIFRelAcc}
\end{eqnarray}
Notice that in this expression the GMLT that transforms the acceleration into terms of the observer $S'_u$ is not  relevant if  the acceleration comes out zero,  as it operates on both parts of the subtraction. This essential feature formally describes the fact that \emph{all} locally coincident LIF's ---i.e. ${S'}^\star$ with whatever velocity ${u'}^\star$ in $\Lambda({u'}^\star, 2)$ --- will ascribe a zero relative acceleration to any coinciding free-falling particle.\\
 Taking into account that parallel transport  occurs with initial and final  frame velocities according to  Eq. (\ref{freefallvelocities}) the transport frame coincides with the free-falling particle. We then immediately see that the remote value $\mathbf{v}_1$ of the velocity  is transformed proportional to  the local value $\mathbf{v}_2$;
\begin{eqnarray}
\Lambda^{-1}(\mathbf{v}_2, 2) \Lambda (\mathbf{v}_1, 1)   \left(\begin{array}{l} 
1   \\
 \mathbf{v}_1  \end{array} \right) &= &\gamma_1^{-1}\gamma_2  \Phi_1 \Phi_2^{-1}  \left(\begin{array}{l} 
1   \\
 \mathbf{v}_2  \end{array} \right)  \label{boost}
\end{eqnarray}
It is clear  that this  transport can be viewed as a boost (time propagator) only in case the initial and final velocities are related by \emph{free-fall acceleration} and as such be identified as ``parallel" transport (which properly relates it to the connections of GRT). This property shows that the transport relation, $\mathbf{1}$ to $\mathbf{2}$, must necessarily be the evolution of \emph{free-fall}; eventual other (unique) evolutions by $\mathbf{a}^* \neq \mathbf{a}$ will not transform $(1, \mathbf{v}_1)$ proportional to $(1, \mathbf{v}_2)$ as in Eq. (\ref{boost}), and will not lead to the cancelation of the relative acceleration Eq. (\ref{LIFRelAcc}). For straightforward  inspection of Eq. (\ref{LIFRelAcc}) shows that the proportionality factor is cancelled in the final fraction and the bracketed term  turns out identically zero, irrespective the velocity of the observer's LIF as we have mentioned above.\\
 The relative acceleration of a free-falling particle in the local coordinates of the observer's LIF-frame ---and all coinciding LIF frames--- is zero:
\begin{eqnarray}
{\mathbf{a}'}_{(PT)}   &=& 0  \label{WEP}
\end{eqnarray}
In the L-P type model, the Weak Equivalence Principle is thus fully satisfied in the LIF perspective:  at the intersection of their free-fall trajectories, the observer and particle have a zero relative acceleration.  This result is of course due to the particular process of parallel transport which is embedded in the calculation of the derivative. While the \emph{standard} derivative  is an isotropic operation, the \emph{covariant} derivative is not; each initial remote value  ---required to make the difference  with the final local value for differentiation--- is rendered local in an anisotropic but unique manner according to  the connectability by free-fall trajectory. Application of  different  transport procedures (or e.g. the standard derivation which implicitly uses invariant transport) would result in residual kinematic acceleration terms  ---as in Eq. (\ref{StatRelAccDiff}) for the static observer--- while with the parallel transport protocol the universality of free-fall emerges, i.e. with independence of the relative velocity. The transport procedure with the invariance of the quantities in the auxiliary free-falling transport frame is thus consistent with the resulting validation of the WEP Eq. (\ref{WEP}).\\
We have in the previous development not invoked the specific form of the free-fall acceleration Eq. (\ref{StaticAcc}). The procedure thus  hinges on the \emph{concept} of free-fall acceleration $\mathbf{a}$ in the definition of the adapted derivative $d/dt_{(PT)}$, according to  Eq. (\ref{freefallvelocities}), but not its explicit form. The \emph{free-fall} evolution however implies the uniqueness condition of the covariant derivative Eq. (\ref{protorelacc}) and the necessary boost condition Eq. (\ref{boost}) for satisfying the WEP.  Thus in principle the WEP in the L-P model can be fully satisfied in the sense of GRT, but in view of the 1-PN order of the free-fall acceleration Eq. (\ref{StaticAcc}) the validation of the WEP in the sense of GRT is only sustained till 1-PN at present.   \\
Concerning the free-fall dynamics of the transporting frame in the case of a free-fall observer, we notice that  an observer submitted to a forced non-LIF motion will again require an adapted definition of the derivative in relation to its forced motion, e.g. in the case of the  \emph{static} observer  the ``adapted'' transport in coordinate space consists in retaining identically the remote value.    The validation of the WEP ---in the sense of  description (b) ---  is the concern for a LIF observer, the description for other types of observers could receive attention in separate dedicated work. \\ 
Finally we remark that in a LIF ---reduced to the intersection point of the orbitals--- the WEP is found satisfied, it is known however that each separation from this point  produces relative acceleration again. We look in the following section how this is described in the L-P type model.

\subsection{Geodesic deviation}
In the LIF  the \emph{tidal} acceleration is proportional to the space and time separation of the particle from the observer. (The procedure of parallel transport includes an irreducible time lag and thus restricts the derivative to time-like separations).   Tidal acceleration in a LIF
 is typically expressed in a configuration in which two separated systems had  identical  velocities ---relative to coordinate space--- at an earlier moment. In GRT the relative acceleration due to  geodesic deviation  along the curve $x^\mu (\tau)$ can be expressed in coordinates of the LIF observer (e.g. \citet{Weinberg1972}, Eq. 6.10.1);
\begin{eqnarray}
\frac{D^2 \delta x^\lambda}{D \tau^2} &=& R^\lambda_{\ \nu \mu \rho}  \delta x^\mu \frac{d x^\nu}{ d \tau}
\frac{d x^\rho}{ d \tau} \label{GRTgeodev}
\end{eqnarray}
where $ \delta x^\lambda$ is the separation  parameter between the particle and the LIF-observer.\\
We calculate the tidal acceleration now according to  the L-P type model. The LIF observer $S'_u$ requires again locally  both the initial and final velocity of the particle at respective instances $\mathbf{1}$ and $\mathbf{2}$ (see Fig. 2). During the present ``observation" the particle remains remote throughout and  thus two remote values  must be rendered local for differentiation.  Hereto auxiliary transporting frames must evolve,  one  from $\mathbf{1}$ to $\mathbf{3}$ followed by $\mathbf{3}$ to $\mathbf{4}$  for the initial value and another, from $\mathbf{2}$ to $\mathbf{4}$  for the final value of the particle's velocity. \\

\hskip 0 cm \includegraphics[width=80mm, height=87mm]{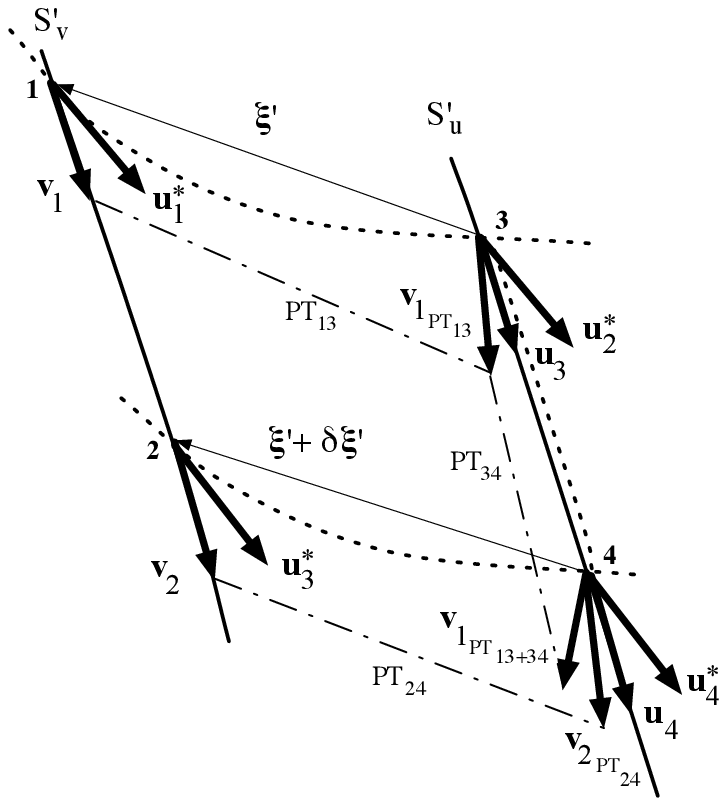}

\vglue - 6.5 true cm
\hangindent = 9 true cm \hangafter =  - 40 {\small \em \noindent  Fig.2:  A LIF observer $S'_u$ with an identical  initial velocity as  a free-falling particle $S'_v$, $\mathbf{u}_3 = \mathbf{v}_1$, separation  by a vector $\mathbf{\xi}$ at space and time instance $\mathbf{3}$, is measuring the latter's relative acceleration.  Auxiliary transport frames  (dotted trajectories) produce the local values $\mathbf{v}_{1_{PT_{13+34}}}$ and $\mathbf{v}_{2_{PT_{24}}}$ at final space and time instance $\mathbf{4}$ of the observer. Notice that the transported initial value of the velocity $\mathbf{v}_{1}$ is obtained by a parallel transport   composed of $PT_{13}$ with initial velocity $\mathbf{u}^\star_{1}$ and final velocity $\mathbf{u}^\star_{3}$ and $PT_{34}$ with initial velocity $\mathbf{u}_{3}$ and final velocity $\mathbf{u}_{4}$. The final part of the parallel transport is thus identical to co-evolution with the observer $S'_u$. 
 \vskip 2.5 cm}

\noindent We introduce shorthand notations now;  in $\Lambda(4)$, $4$ indicates velocity and location $(\mathbf{u}_4, 4)$ and, $4^\star$ indicates $(\mathbf{u}^\star_4, 4)$, etc.\\
The relative acceleration  in the LIF  is then written according to  $\lim_{dt' \to 0} (\mathbf{v}'_{2_{PT_{24}}} - \mathbf{v}'_{1_{PT_{13+34}}})/dt'$, Eq.  (\ref{protorelacc}):
\begin{eqnarray}
 {\mathbf{a}'}^k_{(PT)}  & = &  \lim_{dt' \to 0}
\frac{\Lambda(4)^k_{\ \mu}}{\Lambda(4)^0_{\ \nu}}  \left(
 \frac{\Lambda^{-1}({4^\star})  \Lambda( {2^\star}) ( 1,    \mathbf{v}_2 )^\mu}{\Lambda^{-1}( 4^\star)   \Lambda( 2^\star)( 1,    \mathbf{v}_2 )^\nu }  
-
 \frac{\Lambda^{-1}(4) \Lambda(3) \Lambda^{-1}(3^\star)   \Lambda(1^\star) ( 1,   \mathbf{v}_1 )^\mu}{\Lambda^{-1}(4)  \Lambda(3)\Lambda^{-1}(3^\star)    \Lambda(1^\star) ( 1,    \mathbf{v}_1 )^\nu} 
\right)  \frac{1}{dt'} \label{GMLTgeodev}
\end{eqnarray}
where $dt' = \Lambda (u_4, 4)  ( 1,    \mathbf{u}_4 )^0  dt =  \gamma_4^{-1} (u_4) \Phi_4 dt $. \\
In practice  the expression of the relative acceleration is studied to the first order in the separation four-vector $\mathbf{\xi}^\mu$.  The explicit rendition of the relative acceleration can be done using the expressions for the GMLT  Eqs.(\ref{GMLT},\ref{InverseGMLT}).  It will be clear that in the non-relativistic limit  the usual Newtonian tidal acceleration is recovered  from Eq. (\ref{GMLTgeodev});
\begin{eqnarray}
 {\mathbf{a}'}_{(PT)}   &\approx&    \lim_{dt' \to 0}  (- \delta \mathbf{u}_{43} + \delta \mathbf{v}_{21} + \delta \mathbf{u}^\star_{42} -\delta \mathbf{u}^\star_{31})/dt'   \ = \  {c'} ^2 \nabla' (\mathbf{\xi}.\nabla' \varphi )
\end{eqnarray} 
where we have used, next to configuration settings $\mathbf{u}_3 = \mathbf{v}_1$ and $\mathbf{\xi}_{,\xi_0} = 0$, the standard first-order time developments of the approximated velocities  $\mathbf{u}^\star_1$ and  $\mathbf{u}^\star_2$ of the transport frames;
\begin{eqnarray}
\mathbf{u}_1^\star \approx \mathbf{v}_1 + \mathbf{\xi}/\xi^0 &,& \mathbf{u}_3^\star \approx \mathbf{u}^\star_1 - {c'}^2 \nabla \varphi_1  \xi^0  \ \ \ , \ \ \ \mathbf{u}_2^\star \approx \mathbf{v}_2 + \mathbf{\xi}/\xi^0 \ \ \ , \ \ \  \mathbf{u}_4^\star \approx \mathbf{u}^\star_2 - {c'}^2 \nabla \varphi_2  \xi^0
\end{eqnarray}
and development of the gravitational potential;
\begin{eqnarray}
\varphi_4  \approx  \varphi_2 + \mathbf{\xi}.\nabla \varphi_2 + {\xi_0} \partial_0 \varphi_2  &,& \varphi_2  \approx  \varphi_1 + d \mathbf{x}.\nabla \varphi_1  + dt  \partial_t \varphi_1
\end{eqnarray}
and finally  we have also used the  contravariant space-time GMLT, {$S'$} to {$S_0$} for gradient operators:
\begin{eqnarray}
\nabla &=& {{\bf u}'}   \frac{1}{{\Phi   \left( {\bf r} \right)}} \left( ( {\gamma  \left( u' \right)} - 1)\frac{ {{\bf u}'} . \nabla' }{{u'}^2} +  \frac{1}{{c '}^2}  {\gamma  \left( u' \right)}  \partial_{t'} \right)   + \frac{1}{{\Phi   \left( {\bf r} \right)}} \nabla'  \label{gradientGMLT}  \\ 
\partial_{t} &=& {\gamma  \left( u' \right)} {\Phi   \left( {\bf r} \right)} \left( \partial_{t'} +  {{\bf u}'} .
\nabla'     \right)  
\label{partialderivtGMLT}
\end{eqnarray}
Relativistic corrections to the Newtonian expression can be obtained by developing both the parallel transport and the observer's GMLT.
A parallel transport $PT_{24}$ in coordinate perspective is given to first order ($\delta\varphi_{42}, \delta u^\star_{42}$) by;
\begin{eqnarray}
& & \Lambda^{-1}({\mathbf{u}^\star}_4)   \Lambda({\mathbf{u}^\star}_2) \\
&\approx &\left(\begin{array}{ll}
1- \delta\varphi_{42}    & \delta{\mathbf{u}^\star}_{42} {c}^{-2}  - 2{\mathbf{u}^\star}  {c}^{-2}  \delta \varphi_{42}  \\
\delta{\mathbf{u}^\star}_{42}+   \mathbf{u}^\star (     \delta\varphi_{24}  (2 +      \mathbf{u^*}^2 c^{-2})  + \mathbf{u}^\star.\delta{\mathbf{u}^\star}_{42}   c^{-2} /2 )
&  \mathbf{1}  (1+\delta\varphi_{42} )   + \mathbf{u}^\star_i {\delta \mathbf{u}^\star_{42}}_j c^{-2}/2 - {\delta\mathbf{u}^\star_{42}}_i {\mathbf{u}^\star}_j c^{-2}  /2 \end{array} \right) \nonumber
\end{eqnarray}
This transport expression then needs to work on $ ( 1,    \mathbf{v}_2 )^\mu$. The same expression is required with indices changed $4\to3$, $2\to1$.
The observer's GMLT  at space and time instance $\mathbf{4}$ is approximated  according to ;
\begin{eqnarray}
\Lambda({\mathbf{u} }_4)  &\approx&  \left(\begin{array}{ll}
(1+ {u}_4^2/2 c_4^2) \Phi_4  &  -\mathbf{u}_4   \Phi_4 / c_4^{2} \\
 -\mathbf{u}_4  (1+ {u}_4^2/2 c_4^2) / \Phi_4  &  \mathbf{1} / \Phi_4 + {u_4}_i {u_4}_j/{2 \Phi_4 c_4^2}  \end{array} \right)
\end{eqnarray} 
and all terms in Eq. (\ref{GMLTgeodev}) need to be retained to the required order to match the GRT expression (\ref{GRTgeodev}).\\
 We thus found the process of parallel transport, as defined in subsection (\ref{subsPT}), to be adequately applicable in the definition of the derivative of remote quantities  (\ref{protorelacc}) as shown in the  case of the validity of the WEP and the geodesic deviation in a LIF.

\section{Conclusions.}
In the Lorentz-Poincar\'e type model, the Weak Equivalence Principle  in the perspective of a LIF  was studied by evaluating the relative acceleration of a free-falling particle and an observer at the moment of their coincidence. In this model our analysis of relative acceleration in a LIF  exposed the requirement to evaluate locally a remotely measured value.
Thus in order to define an adapted derivative in a LIF, a unique procedure had to be established to relate these quantities. Transport over a free-fall trajectory uniquely and adequately relates ---in an anisotropic manner--- remote with local values; in the present Lorentz-Poincar\'e type model this relation is formally expressed by means of free-fall correlated gravitationally modified Lorentz transformations.  The GMLT at each instant of the transport relate the invariant local measure ---because consistent with the emergent WEP the local measurement standards appear invariably self-similar over the free-fall transport trajectory---  to the varying measures in coordinate space. 
The GMLT at the initial and the GMLT at the final instance of the free-fall transport trajectory are combined then to  relate the measures  locally and at the remote location. The resulting relation precisely corresponds to parallel transport and the adapted derivative  corresponds to the covariant derivative of GRT.  
In GRT the change of 4-vector $T$  due  to parallel transport  along a  geodesic is expressed through the  connections  $\Gamma^\mu_{\sigma\tau} T^\sigma dx^\sigma$. In the adapted derivative of the  L-P model  the connections implicitly appear as their double contraction over the 4-velocity in the free-fall acceleration.  With the procedure of  \emph{parallel transport}  included in the derivative, the Weak Equivalence Principle as the vanishing of relative free-fall acceleration is then found satisfied in the LIF perspective. Since the GMLT have been shown in previous work to produce 1-PN dynamics of GRT, the validity of the WEP in the L-P type model only corresponds to the same order to the one in GRT. Within the present L-P model itself,  the principle is intrinsically satisfied with respect to the procedure of parallel transport. Finally we have shown that the same procedure of transport can be applied to obtain the geodesic deviation in the L-P type model.

\section*{Acknowledgment}
 An earlier development of the argument was commented by Prof. Saulo Carneiro;   his useful remarks on metric connections are gratefully acknowledged.  Also Prof. Harvey Brown is gratefully acknowledged for discussing  the interpretation of the Equivalence Principle in Relativity Theory. An anonymous referee is acknowledged for pointing out the requirement of necessity and uniqueness of dynamical conditions on the transport procedure and its relation to GRT.\\
This work was supported by FWO--Vlaanderen project F6/15-VC. A87.


\begin{thebibliography}{}


\bibitem[Arminjon (2006)]{Arminjon2006} Arminjon M., Space Isotropy and Weak Equivalence Principle in a Scalar Theory of Gravity, \emph{Brazilian Journal of Physics}, {\bf 36}, 1B, 177-189, 2006


\bibitem[Bell (1987)]{Bell1987} Bell J. S., \emph{Speakable and Unspeakable in Quantum Mechanics}, Cambridge University Press, 1987

\bibitem[Bini \emph{et al.}(1995)]{Bini1995} Bini  D.,  Carini  P and Jantzen R.T., Relative observer kinematics in general relativity, \emph{Classical and Quantum Gravity}, {\bf 12}, 2549-2563, 1995


\bibitem[Broekaert (2005a)]{Broekaert2005a}
Broekaert J.,  A  Lorentz-Poincar\'e-Type Interpretation of  Relativistic Gravitation,   in \emph{ Proceedings of EPS-13: Beyond Einstein -- Physics for the 21st century}, (ESA SP-637), (ed.) M. Cruise. {gr-qc/0510017} 

\bibitem[Broekaert (2005b)]{Broekaert2005b}
Broekaert J.,  On a modified-Lorentz-transformation based gravity model 
 confirming basic GRT experiments, \emph{Foundations of Physics}, {\bf 35}, 5, 839-864, 2005 b. {gr-qc/0309023}.


\bibitem[Broekaert (2004b)]{Broekaert2004b}
Broekaert J.,  A spatially-VSL  gravity model with 1-PN limit of GRT, conference paper \emph{GR17}, 18-23 July 2004, Dublin, (submitted)  2004 b. {gr-qc/0405015} .

\bibitem[Broekaert (2002)]{Broekaert2002}
Broekaert J., Verification of the `essential' GRT experiments in a scalar Lorentz-covariant gravitation, in {\em PIRT VIII proceedings}, (ed.) Duffy M.C., PD Publications, Liverpool, {\bf 1},  37-54, 2002. 

\bibitem[Brown(2005)]{Brown2005} Brown H. R., Physical Relativity, Space-time structure from a dynamical perspective, Oxford University Press (2005).

 

\bibitem[Bunchaft and Carneiro (1998)]{BunchaftCarneiro1998} Bunchaft F., Carneiro S., The static spacetime relative acceleration for the general free fall and its possible experimental test, \emph{Classical and Quantum Gravity}, {\bf 15}, 1557,  1998


\bibitem[Butterfield (2004)]{Butterfield2004}
Butterfield J., On the Persistence of Particles, \emph{Foundations of Physics}, {\bf  35},  233-269, 2004, physics/0401112. 


\bibitem[Cavalleri and Spinelli (1980)]{CavalleriSpinelli1980}
Cavalleri G.,  Spinelli G.,  Field-theoretic approach to gravity in flat space-time, \emph{La Rivista del Nuovo Cimento}, {\bf 3}, 8,1980




\bibitem[Damour (1994)]{Damour1994} Damour T., General relativity and experiment, Proceedings of the XIth International Congress of Mathematical Physics, Paris, July 1994, gr-qc/9412024

\bibitem[Damour (2001)]{Damour2001} Damour T., Questioning the Equivalence Principle,  2001, gr-qc/0109063

\bibitem[Dehnen \emph{et al.} (1960)]{Dehnen1960} Dehnen H., H\"onl H., Westpfahl K., Ein heuristischer Zugang zur allgemeinen Relativit\"atstheorie,  \emph{Annalen der Physik}, {\bf 7}, 6, 370-406, 1960.

\bibitem[Dicke (1957)]{Dicke1957}  
Dicke  R.H., Gravitation without a {P}rinciple of {E}quivalence, \emph{Reviews of Modern Physics}, {\bf 29}, 363-376, 1957. 


\bibitem[Dieks (1987)]{Dieks1987} Dieks D.,  Gravitation as a universal force, \emph{Synthese}, {\bf 73}, 381-397,   1987.











\bibitem[Kenyon (1990)]{Kenyon1990} Kenyon I.R., \emph{General Relativity}, Oxford University Press, 1990


\bibitem[Lorentz (1990)]{Lorentz1900} Lorentz H.A., Consid\'erations sur la Pesanteur, Collected Papers, Vol 5, 198 Martinus Nijhof, The Hague (1937)


\bibitem[McGruder (1982)]{McGruder1982} McGruder III, Ch. H.,  Gravitational Repulsion in the Schwarzschild field,  \emph{Physical Review D}, {\bf 25}, 3191-3194, 1982








 \bibitem[Poincar\'e (1906)]{Poincare1906} Poincar{\'e} H., Sur la dynamique de l'\'electron, \emph{Rendiconti del Circolo Matematico di Palermo}, {\bf 21}, 129-175,1906


\bibitem[Rindler (1979)]{Rindler1979} Rindler W., Essential Relativity. Special, General, and Cosmological (Revised Second Edition) (Springer, Berlin Heidelberg,1979).





\bibitem[Sexl (1970)]{Sexl1970}  Sexl R.U., Universal Conventionalism and Space-Time, \emph{General Relativity and Gravitation}, {\bf 1}, 2,  159-180, 1970.




\bibitem[Stephani (2004)]{Stephani2004} Stephani H., \emph{Relativity. An Introduction to Special an General Relativity (Third edition)},  Cambridge University Press,  Cambridge, 2004



\bibitem[Thirring (1961)]{Thirring1961} Thirring W.E.,  An {A}lternative {A}pproach to the {T}heory of {G}ravitation, \emph{Annals of Physics}, {\bf 16}, 96-117, 1961.

\bibitem[Vlasov (1995)]{Vlasov1995} Vlasov A. A., Phenomenological approach to the viable nonmetric relativistic scalar 4-vector theory of gravitation, \emph{Canadian Journal of Physics}, {\bf 73} 187, 1995 

\bibitem[Weinberg (1972)]{Weinberg1972} Weinberg S., \emph{Gravitation and {C}osmology. {P}rinciples and applications of the {G}eneral {T}heory of {R}elativity},  Wiley, London, 1972.



\bibitem[Will (1993)]{Will1993} Will C.M., \emph{Theory and Experiment in Gravitational Physics}, Cambridge University Press,1995


\bibitem[Wilson (1921)]{Wilson1921}  Wilson  H.A.,  An Electromagnetic Theory of Gravitation, \emph{Physical Review}, {\bf 17}, 54-59, 1921

\bibitem[Winterberg (1988a)]{Winterberg1988a} Winterberg F., Vector Theory of Gravity with Substratum, \emph{Zeitschrift fur Naturforschung}, {\bf 43}, 369-384 (1988)



\end{thebibliography}
\end{document}